\documentclass[sigconf]{acmart}

\usepackage{enumitem}
\usepackage{makecell}   
\usepackage{tcolorbox}  
\usepackage{multirow}
\usepackage{tcolorbox}  
\usepackage{array}  
\newcolumntype{?}{!{\vrule width 1pt}}

\copyrightyear{2022}
\acmYear{2022}
\setcopyright{acmcopyright}\acmConference[MSR '22]{19th International Conference on Mining Software Repositories}{May 23--24, 2022}{Pittsburgh, PA, USA}
\acmBooktitle{19th International Conference on Mining Software Repositories (MSR '22), May 23--24, 2022, Pittsburgh, PA, USA}
\acmPrice{15.00}
\acmDOI{10.1145/3524842.3528446}
\acmISBN{978-1-4503-9303-4/22/05}

\begin{document}

\title{Noisy Label Learning for Security Defects}



\author{Roland Croft}
\affiliation{%
  \institution{CREST - The Centre for Research on Engineering Software Technologies, The University of Adelaide}
  \institution{Cyber Security Cooperative Research Centre}
  \city{Adelaide}
  \country{Australia}
}
\email{roland.croft@adelaide.edu.au}

\author{M. Ali Babar}
\affiliation{%
  \institution{CREST - The Centre for Research on Engineering Software Technologies, The University of Adelaide}
  \institution{Cyber Security Cooperative Research Centre}
  \city{Adelaide}
  \country{Australia}
}
\email{ali.babar@adelaide.edu.au}

\author{Huaming Chen}
\affiliation{%
  \institution{CREST - The Centre for Research on Engineering Software Technologies, The University of Adelaide}
  \institution{Cyber Security Cooperative Research Centre}
  \city{Adelaide}
  \country{Australia}
}
\email{huaming.chen@adelaide.edu.au}

\renewcommand{\shortauthors}{Croft et al.}

\begin{abstract}
Data-driven software engineering processes, such as vulnerability prediction heavily rely on the quality of the data used. In this paper, we observe that it is infeasible to obtain a noise-free security defect dataset in practice. Despite the vulnerable class, the non-vulnerable modules are difficult to be verified and determined as truly exploit free given the limited manual efforts available. It results in uncertainty, introduces labeling noise in the datasets and affects conclusion validity. To address this issue, we propose novel learning methods that are robust to label impurities and can leverage the most from limited label data; noisy label learning. We investigate various noisy label learning methods applied to software vulnerability prediction. Specifically, we propose a two-stage learning method based on noise cleaning to identify and remediate the noisy samples, which improves AUC and recall of baselines by up to 8.9\% and 23.4\%, respectively. Moreover, we discuss several hurdles in terms of achieving a performance upper bound with semi-omniscient knowledge of the label noise. Overall, the experimental results show that learning from noisy labels can be effective for data-driven software and security analytics.
\end{abstract}


\maketitle

\section{Introduction}
\label{sec:introduction}

Mining Software Repositories (MSR) has become a fundamental component of empirical software engineering research, for the purposes of uncovering important findings about software systems and development \cite{hassan2008road}. Particularly, defect datasets have been extensively utilized for the purposes of improving software quality. Such studies either empirically identify correlations between bugs and software/development attributes \cite{menzies2006data, rahman2013how, shihab2010understanding}, or utilize the historical knowledge from software defects to predict future defective modules \cite{hall2011systematic, hanif2021rise, hosseini2017systematic}. These automated processes help solve the limitations of manually assessing large-scale software systems for potential bugs or vulnerabilities. 

Defect data preparation requires code modules to be labeled as clean or defective. To achieve this, researchers typically collect reported post-release software defects for identifying the faulty code modules \cite{croft2021data}. The label correctness is inherently critical for training and evaluation of a prediction model \cite{alonso2015challenges}, and mislabeled instances can heavily influence research outcomes \cite{jimenez2019importance, tantithamthavorn2015impact}. Hence, many studies have made continual improvements to the data preparation process through the investigation of data quality. Noise in bug reports and localization processes can lead to some modules being missed or misclassified as defective \cite{kim2011dealing, herzig2013s, herzig2016impact, fan2019impact}. A more troublesome form of mislabeling arises from dormant bugs \cite{ahluwalia2019snoring, jimenez2019importance}. As practitioners can only utilize defects known to them at the time of data collection, un-detected/reported defects will inevitably be labeled as \textit{clean} modules. 




Practitioners often undergo painstaking efforts to manually verify the entries labeled as defective. However, it cannot be technically applied to the dataset for the verification of the \textit{defective} and \textit{clean} modules as legitimate ones or not in an effective manner, which is critical for the data curation process. It is extremely difficult to ensure the \textit{absence} of vulnerabilities \cite{weinberg2008perfect}, as Software Vulnerabilities (SVs) are security weaknesses in a system, and hence not necessarily functional bugs. Even if the implementation of a design is perfect, it may still be exploitable by an attacker. Furthermore, due to the natural scarcity of SVs amongst the growing size of modern software systems \cite{zimmermann2010searching}, manual verification of clean modules is often infeasible in terms of effort. 

Hence, we assert that perfect ground truth labels are infeasible in practice for SV datasets. These data issues are however expected to have severe negative consequences on data-driven software security tasks, such as Software Vulnerability Prediction (SVP). SVP uses learning-based methods to detect vulnerable modules in code, and is hence highly reliant on the quality of data used to infer the prediction model; \textit{garbage in, garbage out} \cite{domingos2012few}. However, SVP practitioners are inevitably forced to use partial ground truth datasets that include some degree of misclassification \cite{jimenez2019importance}. Hence to overcome this issue, we motivate the need for methods that can handle mislabeled instances. We look towards weak-supervision methods that can better distinguish vulnerabilities despite whether or not there are noises in the dataset.

With the prevalence of label noise and impurities in SV datasets, we aim to leverage the Noisy Label Learning (NLL) methods to learn from these noisy labels. NLL is a research area that has recently gained popularity amongst the Machine Learning (ML) communities \cite{song2020learning}, in response to the commonality of label noise introduced from human annotators and its impacts on the highly expressive nature of Deep Learning (DL) methods. The NLL methods typically aim to cleanse noise from model inputs and outputs~\cite{song2019selfie}, or develop robust, noise tolerant model architecture~\cite{xiao2015learning}. 

In this work, we investigate the effectiveness of NLL techniques for SVP task. Particularly, a two-stage NLL-based method is proposed following a comprehensive experiment design, in which the performance results demonstrate that the method is promising in alleviating the challenge of unavoidable noise in security defect labels. In summary, our main contributions are threefold: 

\begin{itemize}[leftmargin=0.5cm]
    \item We conduct a thorough investigation and characterization of SV label noise and its impacts on prediction models. We provide insight into the instance-dependent nature of SV label noise. 
    \item We are the first to investigate the feasibility of existing NLL methods being adopted to the software engineering domain. Our study reveals that NLL methods exhibit superior performance in comparison with other baselines.
    \item We propose a lightweight approach based on two-stage noise label learning to address the SV label noise issue at model-level. To the best of our knowledge, it is the first work in this direction.
\end{itemize}

The remainder of this paper is organized as follows. Section \ref{sec:background} presents the background knowledge and related work. Section \ref{sec:motivation} contextualizes the motivations behind this study. Section \ref{sec:method} describes our research methodology. We present our findings in Section \ref{sec:results}, and discuss the implications of these findings in Section \ref{sec:discussion}. Finally, we conclude this paper in Section \ref{sec:conclusion}. We have made our dataset and scripts publicly available as a reproduction package from this anonymous link~\cite{reproduction_package}. 

\section{Background and Related Work}
\label{sec:background}

\subsection{Label Noise}
The need for labeled training data is often a significant obstacle to the application of ML models within organizations and industry \cite{figalist2020breaking}. Issues in labeled data can manifest from insufficient time, resources or subject-matter expertise for labeling. These issues produce misleading labels; i.e., label noise. The nature and effectiveness of NLL techniques are dependent on the type of label noise they are targeted against, which can be categorized into two categories~\cite{song2020learning}. 

\textbf{\textit{Instance-independent label noise.}} A naive assumption for label noise is that noisy instances are independent of data features. This is also referred to as symmetric or uniform noise, as each label has the potential to be flipped to another label with equal probability. This can be caused through data corruption or incorrect labeling heuristics. In reality though, such mislabeling is unlikely. 

\textbf{\textit{Instance-dependent label noise.}} A more realistic consideration is that label noise is dependent on both data features and class labels; i.e., certain instances or classes have varying difficulties to label accurately. This instance is also referred to as semantic noise. 

Additionally, there are several assumptions that can be made about the characteristics of the data to assist with noisy label learning. Commonly, we assume the classes of a dataset to be separable (a classifier can perfectly distinguish between any two samples of disparate classes) and to have a smooth distribution (examples that are close to each other in the feature space are more likely to have the same label). 

\subsection{Noisy Label Learning}

Data-Centric AI is becoming of increasing focus within the ML research community due to the growing importance of data. It is well known that \textit{better data beats better algorithms} \cite{domingos2012few}. However, due to prevalence of aforementioned label noise it is highly difficult to obtain ideal datasets. Hence, weakly supervised learning methods aim to overcome data challenges to nevertheless create strong predictive models \cite{zhou2018brief}. 

Noisy Label Learning (NLL) is a form of weak-supervision learning methods that aims to produce effective learning models from noisy datasets~\cite{song2020learning}. Current solutions predominantly fall under two categories~\cite{huang2019o2u}: 1) noise cleaning based approaches that attempt to recognise and cleanse noisy inputs or outputs of a model, and 2) noise-robust models that aim to avoid overfitting to noisy inputs through robust regularization or loss design. 

However, we note that prior work investigating noise-robust models often rely upon particular assumptions around label noise nature \cite{huang2019o2u}. Predominantly, noise-robust methods work best against instance-independent label noise and assume simplistic noise transition matrices. However, these assumptions may be inaccurate for the nature of security defects and latent vulnerabilities, which we further discuss in Section \ref{sec:motivation}. Hence, we focus our investigation into noise cleaning based approaches, which have been shown to perform generally well for instance-dependent label noise \cite{song2020learning}. 

Noise cleaning based approaches attempt to identify and remediate noisy examples in a model's inputs or outputs. Generally, noise cleaning is done prior to model training through \textit{Sample Selection} \cite{song2020learning}, in which we either attempt to identify true or uncertainly labeled examples, from noisy training data. It can best mitigate the impacts on model training via label pruning or refurbishment of uncertain labels. Another alternative however, is to adjust a model's outputs post-training based on prior knowledge or estimation of the transition between correct and noisy labels. This is known as \textit{Noise Adaptation} \cite{song2020learning}.

\subsection{Noise Tolerant approaches for Defect Prediction}
Whilst numerous studies have drawn attention to noise in defect datasets \cite{kim2011dealing, herzig2013s, herzig2016impact, fan2019impact}, the commonly suggested remediation has been for manual data cleaning. Few works have investigated noise tolerant approaches for such data noise. Kim et al. \cite{kim2011dealing} and Catal et al. \cite{catal2011class} both proposed a threshold-based method to detect anomalous class instances for defect prediction. They found these algorithms to marginally improve prediction performance. Pandey and Tripathi \cite{pandey2021empirical} have delivered an empirical study to understand the robustness of different defect prediction classifiers against label noise. While Random Forest was found to be the most robust algorithm, their work focused on using artificially inserted label noise, which failed to resolve the challenges from real-world datasets. Thus, to achieve a better and more practical solution, our study extends this knowledge by investigating the feasibility of a variety of NLL techniques. Furthermore, we are the first to analyse noise tolerance for security defect datasets, which have been suggested to exhibit even greater data quality issues than regular defect datasets \cite{croft2021data}. 

\section{Motivation}
\label{sec:motivation}

\begin{figure}[h]
  \centering
  \includegraphics[width=\linewidth]{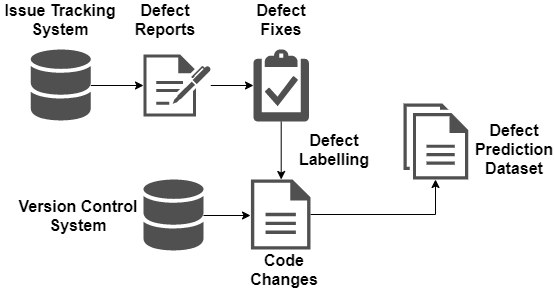}
  \caption{The defect data preparation process.}
  \label{fig:labeling}
  \Description{...}
\end{figure}

Label noise can manifest through a variety of sources for defect datasets, and various studies have demonstrated negative impacts of noise on the performance of defect prediction models \cite{kim2011dealing, herzig2013s, herzig2016impact, fan2019impact}. Defect label noise stems from the reliance of bug reports to identify and localize defects. We provide an overview of the standard defect labeling process in Figure \ref{fig:labeling}. 

Tracing bug reports to code modules can be error-prone due to incomplete information \cite{nguyen2012multi} or tangled code changes \cite{fan2019impact, herbold2020large, herzig2016impact}. However, as detected vulnerabilities are verifiable, it is likely to resolve the issues through extensive manual cleaning. Significant issues arise from sources of noise in the negative class (i.e., non-defective modules), as bug reports only reveal known and detected defects. Hence, dormant, latent or undetected defects that exist within the source code are inevitably contained in the negative class of a dataset. Prior authors have observed the impact of mislabeled latent defects \cite{jimenez2019importance,falessi2021impact}. They have recommended to remediate this issue by using more complete defect data at a later time period, when more defects have been detected. However, this may not always be feasible as models may need to be constructed straight away for continuous bug prediction \cite{wang2021continuous}. Zheng et al. \cite{zheng2021d2a} recently proposed the D2A dataset that used static analysis and manual verification for additional labeling indicators. Whilst these efforts can help to uncover latent vulnerabilities in a more timely way, it is still limited to \textit{known} vulnerabilities. A practitioner would only ever be able to train a predictive model on a \textit{partial} ground truth dataset that would include some degree of misclassification. 

For SVs, practitioners additionally require bug reports to be labeled as security relevant, but this additional labeling is incomplete and unreliable in practice. Security defects may not be completely labeled as such in a bug tracking system. Jiang et al. \cite{zhoufinding} found that over 65\% of vulnerability patches were \textit{silently} reported. Through traditional data curation process, these silent vulnerabilities would also be incorrectly labeled in an SV dataset. 

To investigate such noise, we have manually curated vulnerability labels for Mozilla Firefox source code files. We firstly collected reported vulnerabilities for a relevant release (i.e., SVs detected during development of the next release). For latent vulnerabilities, we examined SVs detected in future releases and traced their existence in the source code to determine affected prior releases. This gives us an indication of latent vulnerabilities, but it is by no means complete knowledge as we are still relying on \textit{detected} vulnerabilities from an extended time period. However, despite this incomplete knowledge, we observed over three times as many latent vulnerabilities than the available reported vulnerabilities. Hence, label noise is a significant factor in SV datasets. 

We note that amongst other difficulties, data labeling for security defect datasets is extremely effort intensive. In this study, we have manually curated a dataset from three years of Mozilla Firefox development (the previous 22 releases). Manual verification of the reported and latent vulnerabilities consumed over 100 hours of effort from two experienced researchers in this area. Hence, data cleaning may even be infeasible due to effort requirements for large-scale datasets. Modern software systems contain millions of lines of code. 

\begin{figure*}[h]
  \centering
  \includegraphics[width=\linewidth]{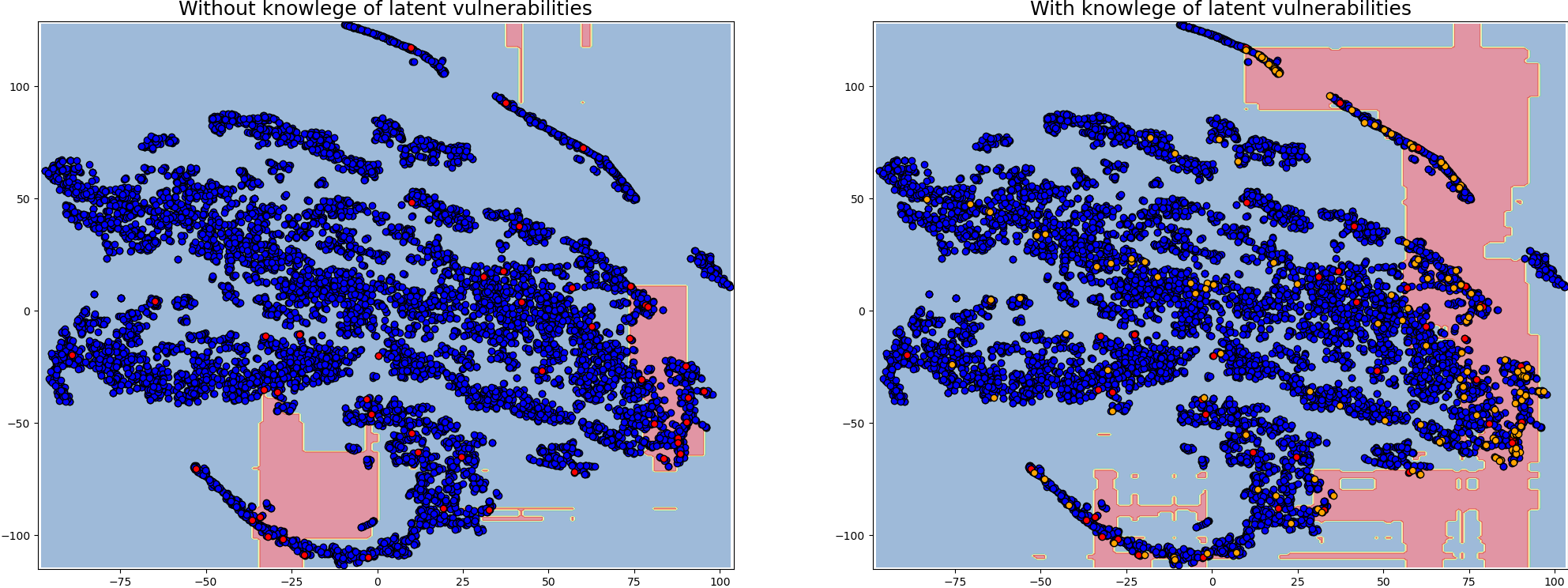}
  \caption{The decision surface inferred from a prediction model without and with knowledge of latent vulnerabilities. Red dots indicate reported vulnerabilities, orange dots indicate latent vulnerabilities reported during an additional two year time period, and blue dots indicate suspected clean modules. The red and blue areas display the decision boundary of a random forest classifier for positive and negative predictions, respectively.}
  \label{fig:motivation}
  \Description{...}
\end{figure*}

Figure \ref{fig:motivation} displays two software vulnerability prediction models trained from standard software metric features \cite{theisen2020better} from the source code of Mozilla Firefox Release 65. However, we stress that the embeddings and decision boundaries displayed in Figure \ref{fig:motivation} are just an approximation, as the features have been embedded using t-SNE \cite{van2008visualizing}. The graph on the left displays the known vulnerabilities at training time (i.e., for Release 66), and the figure on the right displays the known vulnerabilities from an additional two years of latent vulnerability detection. We can see that knowledge of additional labels significantly alters the produced decision boundary of a classifier. Furthermore, without knowledge of the latent vulnerabilities, the model fails to predict many of them. 

We assert that it is infeasible in practice to obtain complete knowledge of all exploitable SVs in a codebase at any given time. Hence, label noise is a persistent issue for SVP and security defect analysis. We motivate the need for model-level solutions to such data challenges, to produce methods that are still able to learn and form strong predictors from noisily labeled SVs. 

From the observable labels in Figure \ref{fig:motivation}, we also conjecture SV noise to be label-dependent (i.e., semantic noise). The latent vulnerabilities appear to be non-randomly distributed within the source code files, and hence label impurities are likely unable to be solved with a simple noise transition matrix or improved robustness. SV label noise is also heavily skewed towards the negative class. 

\section{Research Methodology}
\label{sec:method}
In this section we describe the experimental setup used to achieve our research aims; to investigate the capabilities of NLL methods for improving SVP robustness. We ask three main Research Questions (RQs) to help guide our analysis. 

\textbf{RQ1. What are the vulnerability types, latencies and locations of noisily labeled files for the Mozilla dataset?} We first attempt to characterize the label noise in our examined dataset. Such knowledge can help in detecting future label noise, and also serve to identify the strengths and weaknesses of the NLL models in our subsequent RQs. 

\textbf{RQ2. How effective are NLL techniques for inferring label noise in SV training data?} We investigate the capability of NLL techniques to accurately identify noisy samples that we are aware of in the training datasets. This investigation provides better insights into the proficiency of the NLL methods and their extent to clean data samples prior to model training. We also aim to determine through this RQ whether we can use NLL for identifying vulnerabilities in current releases via automated noise cleaning, rather than in future releases via predictive modeling. 

\textbf{RQ3. How effective are NLL techniques for SV prediction?} We investigate the capability of NLL techniques to assist SVP models to predict the presence of vulnerabilities in future releases. These insights help us to determine the extent to which we can improve the effectiveness and reliability of SVP models through noisy label learning, and hence produce better methods for automatic SV detection, and use of SV datasets.

\subsection{Data Collection}
Whilst researchers have created and shared several SV datasets \cite{zhou2019devign, chakraborty2021deep, fan2020ac}, we opted to create a new dataset in order to gain more knowledge and confidence of the labeling process. We selected the Mozilla Firefox project for analysis for several reasons. Firstly, it is one of the largest open-source web browsers with over 200 million active users as of November 2021\footnote{\url{https://data.firefox.com/dashboard/user-activity}}. Secondly, it has over a decade of active and publicly available development history. Thirdly, it maintains thorough and public bug reporting practices \cite{mozillaadvisory}. Finally, it has been the subject of many prior SV research studies~\cite{theisen2020better, yu2019improving}. 

Mozilla Firefox periodically updates its version via releases. We downloaded the source code of each public Firefox release. SVs that affected prior releases are publicly disclosed to the Mozilla Security Advisory \cite{mozillaadvisory} upon each new release. To assign the vulnerability labels in our dataset, we obtained the relevant bug report for each advisory entry, and then manually identified the vulnerable files in the prior release through examination of the vulnerability fix. This manual verification was conducted by two individuals to increase the reliability and reduce potential bias; the first author and a postgraduate software engineering student. The two raters achieved a Cohen's Kappa value of 0.701 \cite{cohen1960}, which implies substantial agreement. Disagreements were then resolved through discussion. Due to the intensive effort required for this manual verification process (approximately 100 hours), we obtained labels for the past three years of Mozilla Firefox development. This time period covers 22 releases (release 63-84).  Each release had a mean of 9619 files and 20 reported SVs, as displayed in Table \ref{tab:data_stats}. In total, we obtain 211,658 files and 443 reported SVs. 

Additionally, to aid our evaluation of NLL techniques, we attempted to identify noisy labels within the non-vulnerable code. However, as stated in our earlier assertions, it is infeasible for us to obtain a perfect ground truth set of labels; we cannot accurately identify all undetected or unreported SVs. However, we can identify some \textit{partial} sources of noise, to provide a general indication. 

We firstly identified latent vulnerabilities, using knowledge from future version fixes. As future knowledge is required to identify these SVs, they would be undetectable at model training time \cite{jimenez2019importance}. For each disclosed SV in the Mozilla Advisory, we identified the versions prior to the affected release in which the vulnerable code was still present. To do so, we only considered files as containing a latent vulnerability if they contained all original code lines deleted or modified by a vulnerability fix; i.e., the vulnerable lines of code. Fixes that only contained line additions (i.e., through an added check) were unable to be reliably verified in prior versions \cite{le2021deepcva}, and hence remained as unknown latent SVs in our dataset. Although sources like NVD contain information about affected versions, this information is unusable as it is unreliable \cite{anwar2021cleaning}. Hence, we needed to identify these labels manually. 

Secondly, we attempted to identify unreported or \textit{silently} patched SVs \cite{shu2019better}. We considered these as any security related defect that has been fixed in the bug reporting system, but not disclosed via the security advisory. As each release contains thousands of bug reports, it was infeasible to manually scan every bug fix. Hence, we used a keyword search approach to narrow our search space, similar to Theisen et al. \cite{theisen2020better} and Yu et al. \cite{yu2019improving} in their data collection approaches of the Mozilla Firefox dataset. We decided to extract keywords from the Common Weakness Enumeration (CWE) Top 25 vulnerability list\footnote{\url{https://cwe.mitre.org/top25/archive/2021/2021_cwe_top25.html}}; as they are the 25 most prevalent SV types. Our keyword list can be found within our reproduction package \cite{reproduction_package}. The same two individuals as before then manually verified the security relevance of each bug report that contained one of these security keywords. A bug report was considered as security relevant if its description aligned with an existing CWE type. We also conducted manual verification of the presence of these security defects in the relevant releases by checking the source code. 

Table \ref{tab:data_stats} displays the minimum, mean, maximum and total number of source code files and vulnerabilities in the 22 Mozilla Firefox releases. We observe our dataset to be extremely class imbalanced when using the common assumption that files without detected SVs are non-vulnerable. There is a considerably larger number of noisy SVs (Silent and Latent) than reported SVs. 

\begin{table}[tb]
  \caption{Data statistics for Mozilla Firefox release 63 to 84.}
  \label{tab:data_stats}

  \begin{tabular}{c|c|ccc}
    \hline
     & \multirow{2}{*}{\textbf{Files}} & \multicolumn{3}{c}{\textbf{Vulnerabilities}}\\
     &  & \textbf{Reported} & \textbf{Silent} & \textbf{Latent}\\
    \hline
    
    Min & 9354 & 0 & 0 & 0\\
    Mean & 9621 & 20 & 22 & 61\\
    Max & 9849 & 54 & 48 & 152\\
    \hline
    Total & 211,658 & 443 & 474 & 1341\\

    \hline
\end{tabular}
\end{table}

To inspect the label noise characteristics for RQ1, we examined the CWE type of each SV. CWE provides a hierarchical classification scheme for vulnerabilities \cite{CWE}. To reduce the dimensionality of our analysis, we combined similar CWE categories into higher-level categories using the hierarchical structure of CWE \cite{paul2021security}.

\subsection{Software Vulnerability Prediction}
Software Vulnerability Prediction (SVP) is the process of using a learning-based model to automatically predict or detect the presence of an SV in a given module. SVP has been extensively researched by software security researchers \cite{hanif2021rise}. It is a data-intensive task. To construct SVP models, we followed the standard guidelines from prior literature \cite{ghaffarian2017software}.

We firstly opted to conduct our classification at the file-level; we assigned each source code file a label as to whether it contains a reported vulnerability or not. The majority of SVP research has been conducted at the file-level \cite{croft2021data}, but recent state-of-the-art models have moved towards finer granularities \cite{lin2020software}. However, we choose to retain our prediction at the file-level, as further localisation of vulnerable code within files would introduce additional noise and distrust in our positive labels. For instance, tangled commits can make function or commit level localisation error-prone \cite{herzig2016impact}. Without confidence in our positive labels, our evaluation would be unreliable, which is unsuitable for an investigative study such as ours. We only retained .c or .cpp files. Additionally, we removed any file containing the word \textit{test} in the file path, or third-party packages that were bundled with releases, contained in the \textit{third\_party} directory. We removed all comments from each file. 

To encode file-level features, we tokenized and embedded the source code of each file using CodeBERT \cite{feng2020codebert}; a state of the art code embedding model based on the RoBERTa architecture \cite{liu2019roberta}, which has been trained on millions of programming language examples. We selected CodeBERT embeddings due to their prominence in recent literature, promising performance in this domain \cite{pan2021empirical}, and capabilities for enhancing the use of small sized datasets \cite{prenner2021making}. We used a random forest classifier to perform classification, based on its proven success for file-level prediction in prior works \cite{jimenez2016vulnerability}. We did not evaluate any DL-based classifiers, as they are not as suitable for file-level prediction due to their data hungriness~\cite{xin2018machine}. 

\subsection{Baselines}
For our baselines, we constructed SVP models using the described setup in Section 4.2. A side effect of label noise in SV datasets is that it perpetuates extreme class imbalance, due to a significant number of vulnerable cases being hidden in the negative set. This is demonstrated in Table \ref{tab:data_stats}, as we can see that there is a much larger average number of noisy labels. Hence, to mitigate this issue, we adjusted the class weight of each classifier to be balanced. The weight of each sample is made inversely proportional to its class frequency, which is estimated through the number of observed samples in the training set. We also investigated baselines that use class rebalancing to make the number of samples in the training set for each class equal. We only investigated oversampling, due to the extremely low number of samples in the positive class \cite{jimenez2019importance}. We experimented with using both Random Over Sampling and Synthetic Minority Oversampling Technique (SMOTE) \cite{chawla2002smote}. However, we only report the results of Random Over Sampling as we observed it to have better performance than SMOTE when used on the baseline model. 

As an additional point of comparison, we also considered the performance of a baseline model that has the knowledge of the additional noisy labels we extracted during the data collection process (latent and silently fixed SVs). Although a model would not traditionally have the knowledge of these labels at training time, we viewed knowledge of these partially identified sources as an ideal case \cite{jimenez2019importance}. Hence, we investigated the performance difference under this setup, and denote it as the \textit{Semi-Optimal} scenario. 

Considering our assertion towards the unreliability of the negative class, we also considered one-class classification (or one-class learning) as an appropriate benchmark. One class classification learns from a training set that exclusively contains samples of a single class \cite{tax2002one}. We trained and tuned a one class SVM model only using the reported SVs of each release as an additional baseline. 

\subsection{Noisy Label Learning Technique Selection}
We aim to utilise noise cleaning based NLL techniques to help SVP models identify the incorrectly labeled SVs  during learning. To achieve semantic noise cleaning, NLL techniques often use model-based methods to determine confident and uncertain labels \cite{song2020learning}. The most common method to achieve this is via loss analysis; noise-free examples mostly generate small loss \cite{huang2019o2u}. However, for class imbalanced problems such as SVP, the model converges to major classes faster than minor classes. Consequently, the minor class of focus exhibits the largest loss, making these techniques unsuitable for our problem. 

Alternatively, we can indicate selected clean examples to a model to help NLL techniques to learn to differentiate between clean and noisy labels. For SVs, we mainly have confidence in the vulnerable class labels as we can verify their presence from vulnerability advisories and bug reports. However, we have less confidence in the labels for the negative class. In this sense, we treat the positive class as our clean labeled examples, and utilize the remaining examples as unlabeled noisy data. Using this heuristic, we selected appropriate NLL noise cleaning approaches based on Positive and Unlabeled (PU) learning algorithms \cite{bekker2020learning}. Our selected NLL techniques are summarized in Table \ref{tab:techniques}. We investigated two variants for each sample selection technique: pruning the identified noise, or adjusting the labels of the identified noise. 

\begin{table}[tb]
  \caption{Selected NLL techniques for analysis.}
  \label{tab:techniques}

  \begin{tabular}{cc}
    \hline
    \textbf{Technique} & \textbf{Type}\\
    \hline
    
    Confident Learning \cite{northcutt2021confident} & Sample Selection\\
    Multi-Stage Learning \cite{bekker2020learning} & Sample Selection\\
    Post-Processing \cite{elkan2008learning} & Noise Adaptation\\

    \hline
\end{tabular}
\end{table}

\textbf{\textit{Confident Learning}} \cite{northcutt2021confident} first estimates the class prior probabilities for non-vulnerable samples through k-folds validation. The expected label noise is then estimated by examining the number of non-vulnerable samples that have a predicted probability greater than the average predicted probability of the vulnerable class. Non-confident labels are then identified as the non-vulnerable samples with the greatest difference in predicted probability to the average prediction probability of the non-vulnerable class. The number of non-confident samples selected is based upon the expected amount of label noise. We used the implementation provided by Northcutt et al. \cite{northcutt2021confident}. In the standard case, all non-confident labels are removed from training. However, we also consider a variant in which all non-confident labels are flipped to the positive class. We denote this variant as \textit{Confident Learning +}. 

\textbf{\textit{Multi-Stage Learning}} attempts to identify samples that we are confident are non-vulnerable in the training dataset; reliable negatives. We assume that file embeddings of the same class stay in close proximity. Hence, reliable negatives are files in the unlabeled set that are as \textit{different} as possible from vulnerable files. We identified reliable negatives using a nearest centroid approach \cite{le2020puminer}, although other approaches have been proposed \cite{bekker2020learning}. We obtained the centroid of the vulnerable and unlabeled sets. Assuming that the majority of the unlabeled files are non-vulnerable, then the unlabeled centroid would represent the negative class more than the positive one. We calculated the cosine distance of each file to each centroid. We selected the cosine distance as it is a good measure for semantic embeddings \cite{le2020puminer}. Files that were closer to the unlabeled centroid than the positive centroid were selected as reliable negatives. For the simple pruning variant, we removed all files that are not classed as vulnerable or reliable negatives. We refer to this as \textit{1-Stage Learning}. However, for the label adjustment variant, we used the reliable negatives and vulnerable samples to train a random forest classifier that labels the remaining unlabeled posts. We denote this variant as \textit{2-Stage Learning}.  

\textbf{\textit{Post-Processing}} adopts the PU Learning algorithm proposed by Elkan and Noto \cite{elkan2008learning}. This method uses an additional holdout set that is taken from the training data, to determine the average prediction probability of the vulnerable class. Model outputs are then altered to better match the expected distribution, by dividing each prediction probability with the average probability of the vulnerable class.

\subsection{Evaluation Setup}
\begin{figure}[h]
  \centering
  \includegraphics[width=0.8\linewidth]{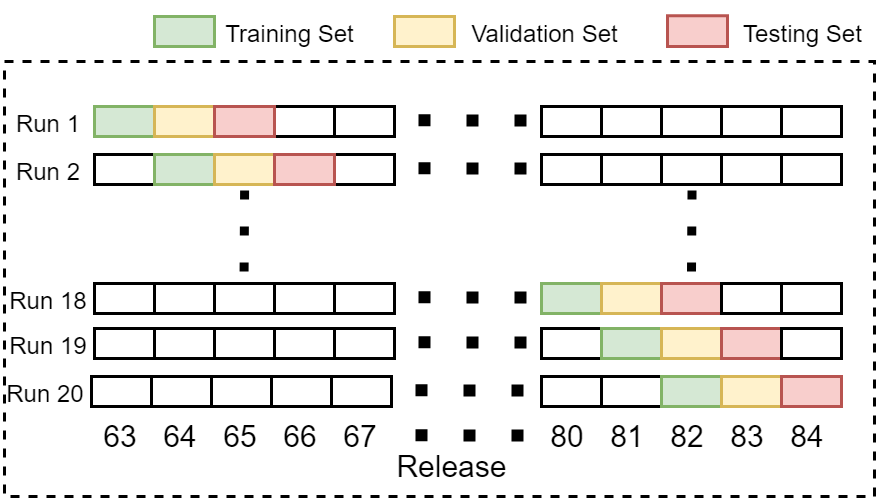}
  \caption{The next release validation setup.}
  \label{fig:next_release}
  \Description{...}
\end{figure}

To evaluate our methods we used next release validation \cite{jimenez2019importance, wang2021continuous}. For each test release x, release x-1 is used for validation, and release x-2 is used for training. We only used training data from a single release, in comparison to data from all prior releases, as it has been shown to result in better performance due to the temporal nature of this data \cite{harman2014less,hovsepyan2016newer}. Figure \ref{fig:next_release} illustrates this process. For RQ2, we assessed the identified noisy labels from the NLL techniques only within the training release. 

The appropriate hyperparameters were tuned on the validation release for each model using Bayesian Optimisation \cite{bayesopt}. We also tuned each NLL technique using the validation release. For confident learning, we tuned the ratio of samples to identify as non-confident. For multi-stage learning, we tuned the distance thresholds for determining reliable negatives. To save space, all hyperparameter values are available in our reproduction package~\cite{reproduction_package}. The NLL techniques and baselines are evaluated on the test release. 

We considered multiple evaluation metrics to measure performance. We primarily considered the Area Under the receiver operator characteristic Curve (AUC) as it has been recommended by other researchers \cite{tantithamthavorn2018experience}. Default classification thresholds of 0.5 are unsuitable due to the imbalanced nature of our dataset. Hence, threshold-independent measures, such as AUC, provide a better indication of performance \cite{tantithamthavorn2018experience}. AUC was used for model tuning and hyperparameter selection on the validation set. For threshold-dependent measures, we determine an optimal classification threshold using Youden's J statistic \cite{youden1950index} on the AUC, which aims to optimize recall whilst minimizing the false positive rate. 

We use AUC as an overall metric to consider both model precision and recall. However, following our prior assertions, we do not have reliable knowledge of the negative labels in our dataset. False positives may actually be mislabeled instances, which would cause related measures to report much lower estimations than their true value. Hence, we also focus our evaluation on the recall of our methods, which is calculated only using our verified true vulnerabilities. We still document precision for completeness however, despite it's potential inaccuracy. For PU learning, a reliable geometric mean of the recall and precision can be inferred \cite{bekker2020learning}. Whilst the value of this metric does not hold any meaning on its own, it achieves a high value when precision and recall would be high, and vice versa. 

\section{Results and Analysis}
\label{sec:results}

\subsection{RQ1: What are the vulnerability types, latencies and locations of noisily labeled files for the Mozilla dataset?}
Firstly, we considered the types of vulnerabilities for the labeled and noisy SVs that we have manually observed in our dataset. Table \ref{tab:cwe_types} displays the CWE types present in our dataset, for each type with more than 10 separate cases. We note that the dataset is heavily skewed towards buffer overflow type vulnerabilities. 

\begin{table}[tb]
  \caption{CWE type frequency of the Mozilla dataset, and the average latency in days.}
  \label{tab:cwe_types}

  \begin{tabular}{cccc}
    \hline
    \textbf{CWE-ID} & \textbf{Definition} & \textbf{Frequency} & \textbf{Latency}\\
    \hline
    
    CWE-119 & Buffer Overflow & 242 & 209\\
    CWE-416 & Use After Free & 38 & 246\\
    CWE-200 & Information Exposure & 31 & 113\\
    CWE-20 & Improper Input Validation & 19 & 26\\
    CWE-79 & Cross Site Scripting (XSS) & 16 & 169\\
    CWE-362 & Race Condition & 14 & 131\\
    CWE-287 & Improper Authentication & 13 & 229\\
    CWE-346 & Origin Validation Error & 10 & 33\\

    \hline
\end{tabular}
\end{table}

Latent vulnerabilities can persist for varying amounts of time, which we define as the latency. We recorded the latency (in days) of each known latent vulnerability. Using a Kruskal-Wallis H Test \cite{kruskal1952use}, we observed the latency to significantly differ for different CWE types at a 5\% significance level ($H=17.344, p=0.015$). Hence, different SV types have varying latency, and varying noise. Table \ref{tab:cwe_types} displays the mean number of days that SVs in our dataset remain latent. 

Buffer Overflow, Use After Free, and Improper Authentication exhibit the longest average latency periods. Buffer Overflow and Use After Free are also the most common types of SV, which alludes to the high amount of noise present in SV datasets. Conversely, Improper Input Validation and Origin Validation Error vulnerabilities exhibit the shortest average latency periods. Hence, these SVs may be more easy to detect by developers. 

\begin{figure}[h]
  \centering
  \includegraphics[width=0.9\linewidth]{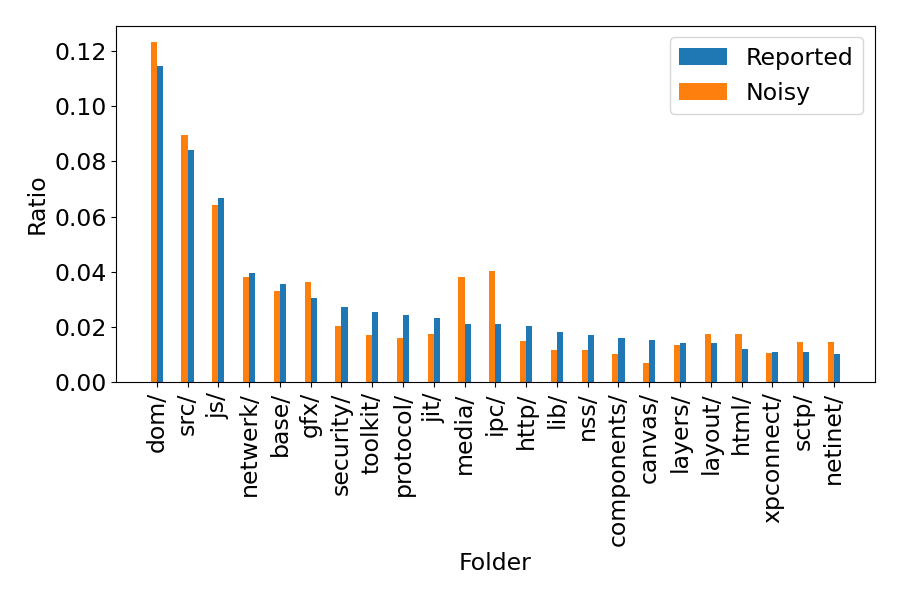}
  \caption{Proportion of source folder locations for reported and latent/noisy vulnerabilities in Mozilla Firefox.}
  \label{fig:file_types}
  \Description{...}
\end{figure}

We also considered the functionality of the files that may receive noisy labels. We tokenized the file paths of the files containing reported vulnerabilities, and noisily reported vulnerabilities. We again only considered tokens that appear more than 10 times in each set. Figure \ref{fig:file_types} displays the proportion of reported or vulnerable files for each Mozilla source folder. 

We confirmed a statistically significant difference in proportions of file locations between reported and noisy labels at a 95\% confidence level, using a chi-square goodness of fit test ($X^2=75.528, p<0.001$) \cite{pearson1900x}. Particularly, we observed a higher proportion of noisy labels located in the \texttt{media/} and \texttt{ipc/} (inter-process communication) folders. Upon manual inspection, we discovered some vulnerable files in these folders exhibited very high latency periods. 

\subsection{RQ2: How effective are NLL techniques for inferring label noise in SV training data?}
\begin{figure*}[h]
  \centering
  \includegraphics[width=\linewidth]{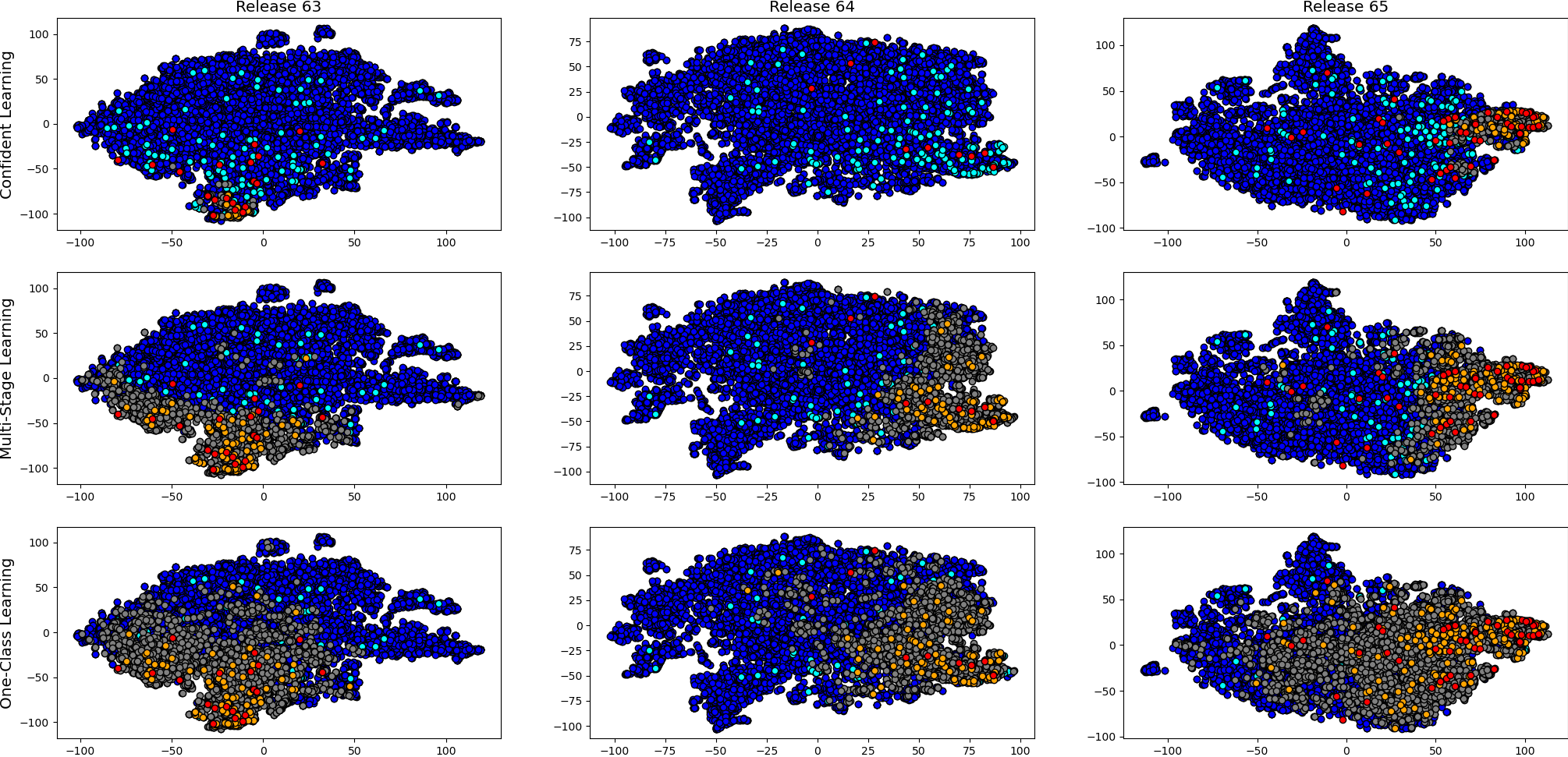}
  \caption{Reliable negatives identified by NLL methods. Embeddings are encoded using t-SNE. Blue = identified reliable negatives. Cyan = vulnerable files identified as reliable negatives (i.e., incorrect). Grey = suspected vulnerable files. Orange = manually identified latent or silent vulnerabilities. Red = reported vulnerabilities.}
  \label{fig:embeddings}
  \Description{...}
\end{figure*}

We investigated the capabilities of each selected NLL technique to identify label noise within the unlabeled set. Figure \ref{fig:embeddings} displays the identified reliable negatives of each method for the first three Mozilla Firefox releases in our dataset, where a reliable negative is a non-vulnerable file that is considered to not have label noise. An optimal technique would be able to identify most of the latent or silent vulnerabilities as suspected vulnerable files, whilst minimizing the number of vulnerable files identified as reliable negatives. We included one-class learning as a baseline comparison.

Using our manually identified set of noisy labels, we evaluated each approach's capabilities for classifying vulnerable files in the unlabeled set in terms of precision, recall and geometric mean. We treated the reliable negatives identified by a technique as non-vulnerable classifications, and the suspected noisy labels as vulnerable classifications. Table \ref{tab:trans_results} displays these performance values. 

\begin{table}[tb]
  \caption{Mean performance for each approach for classifying noisy labels within the training set. Highest values are in bold.}
  \label{tab:trans_results}

  \begin{tabular}{c|ccc}
    \hline
    \textbf{Technique} & \textbf{Precision} & \textbf{Recall} & \textbf{Gmean}\\
    \hline
    
    Confident Learning & \textbf{0.041} & 0.113 & 3.553\\
    Multi-Stage Learning & 0.029 & \textbf{0.555} & \textbf{52.429}\\
    One-Class Learning & 0.020 & 0.454 & 42.438\\

    \hline
\end{tabular}
\end{table}

We observed that one-class learning has very poor precision and tends to over classify samples as vulnerable. This method relies too heavily on the reported vulnerabilities to classify the unlabeled points. It also relies too heavily on the separability assumption, as it assumes the positive and negative classes to not overlap. Conversely, the higher precision value for confident learning indicates that it could yield higher-fidelity suspected vulnerable files while reporting less false positive. However, for false negatives, which is the \textit{Cyan} in Figure \ref{fig:embeddings}, it has a higher number shown with a lower recall value. This results in even lower overall performance than one-class learning. Multi-stage learning appears to have the strongest heuristic for noise identification. It outperforms both one class learning and confident learning in terms of recall and geometric mean. 


However, we observe the techniques to exhibit poor precision towards identifying noisy labels in the training set. Hence, we suggest that these approaches are not yet usable solely for the purposes of noise identification and cleaning. However, the additional knowledge gained from the noisy labels may help classifiers improve their learning of the vulnerable class. This spurs our investigation for RQ3 to see if NLL techniques can improve SVP performance. 

\subsection{RQ3: How effective are NLL techniques for SV prediction?}


\begin{figure}[h]
  \centering
  \includegraphics[width=\linewidth]{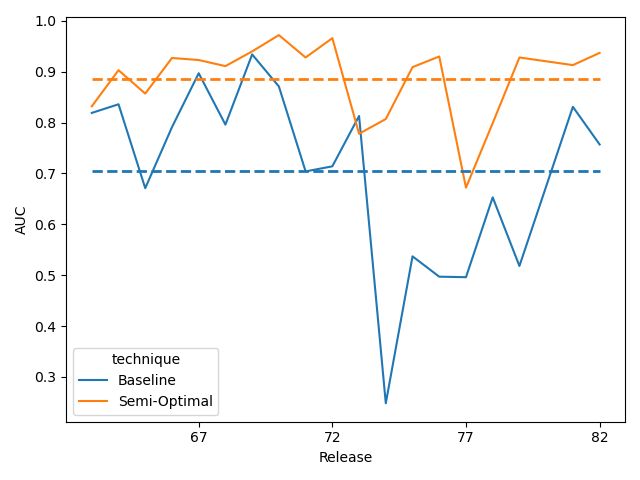}
  \caption{Comparison of baseline model performance to semi-optimal knowledge of noisy labels. Dashed lines indicate mean performance over all releases.}
  \label{fig:baseline_knowledge}
  \Description{...}
\end{figure}

Mean baseline AUC scores range from 0.7-0.8, which is on par with existing studies that use a next release validation setup \cite{jimenez2019importance, wang2021continuous}. Figure \ref{fig:baseline_knowledge} displays the AUC performance of our baseline model in comparison to when we have semi-optimal knowledge of label noise. We observed the performance to significantly increase for nearly every release, which demonstrates the inaccuracy of assuming the unlabeled set as entirely negative examples.  

\begin{figure*}[h]
  \centering
  \includegraphics[width=\linewidth]{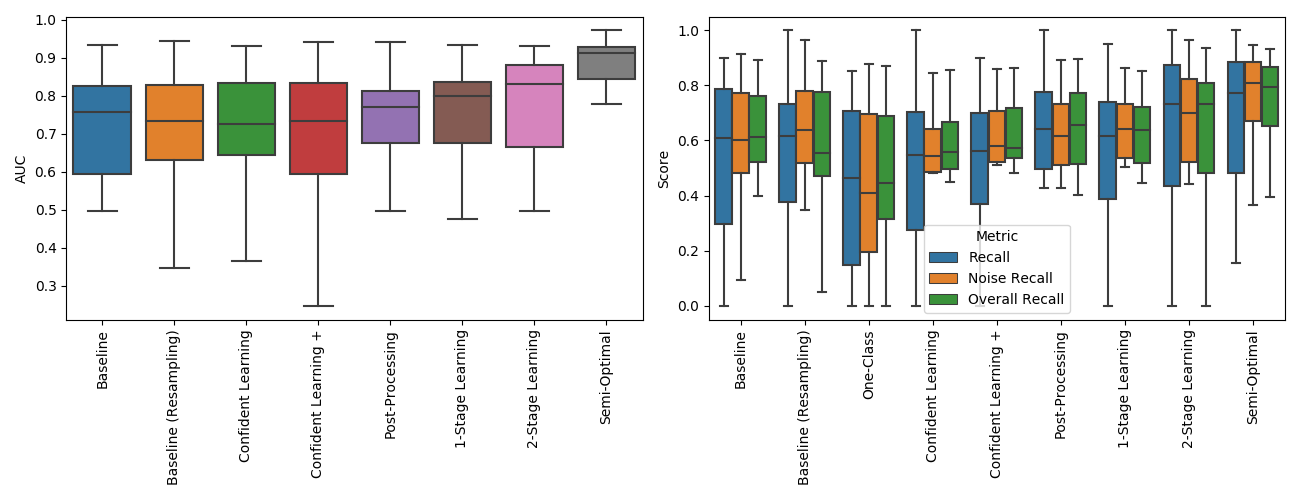}
  \caption{Next release validation performance for baseline and noisy label learning techniques.}
  \label{fig:inductive_results}
  \Description{...}
\end{figure*}

\begin{table*}[tb]
  \caption{Mean performance for next release validation of each prediction technique. Recall is calculated using the reported SVs, the noisy SVs (latent or silent), and overall.}
  \label{tab:inductive_results}
  \begin{tabular}{p{3.5cm}|p{2.5cm}|p{1.75cm}p{1.75cm}p{1.75cm}|p{2.5cm}}
    \hline
     \multirow{2}{*}{\textbf{Technique}} & \multirow{2}{*}{\textbf{AUC}} & \multicolumn{3}{c|}{\textbf{Recall}} & \multirow{2}{*}{\textbf{Overall Precision}}\\
     &  & \textbf{Reported} & \textbf{Noisy} & \textbf{Overall} & \\
    \hline
    
    Baseline & 0.704 & 0.534 & 0.557 & 0.578 & 0.027\\
    Baseline (Resampling) & 0.687 & 0.521 & 0.576 & 0.577 & 0.026\\
    One-Class & - & 0.448 & 0.420 & 0.457 & 0.022\\
    \hline
    Confident Learning & 0.717 & 0.507 & 0.500 & 0.534 & 0.030\\
    Confident Learning + & 0.704 & 0.519 & 0.533 & 0.555 & 0.030\\
    Post-Processing & 0.733 & 0.606 & 0.566 & 0.607 & 0.029\\
    1-Stage Learning & 0.734 & 0.532 & 0.572 & 0.594 & 0.032\\
    2-Stage Learning & 0.773 & 0.659 & 0.633 & 0.665 & 0.030\\
    \hline
    Semi-Optimal & 0.886 & 0.688 & 0.709 & 0.730 & 0.100\\

    \hline
\end{tabular}
\end{table*}

Table \ref{tab:inductive_results} and Figure \ref{fig:inductive_results} illustrate the relative performance of each technique. We evaluated recall for reported vulnerabilities, known noisy vulnerabilities, and overall. We also recorded the overall precision for completeness, but it is not the focus of this study. We observed mixed success for SVP using NLL techniques. Some techniques were able to increase the baseline performance. However, using a Wilcoxon Signed-Rank test \cite{wilcoxon1992} with a 95\% confidence level, we only observe AUC to be significantly higher than the baseline for 1-stage learning ($W=146.5, p=0.020$) and 2-stage learning ($W=154, p=0.001$). Furthermore, none of the techniques were able to produce performance that reached the semi-optimal scenario.

Out of the investigated NLL techniques, 2-stage learning performed the best in terms of both AUC and recall. We observed a mean increase in AUC of 8.9\%, and an increase in overall recall of 15.1\% (23.4\% for the reported SVs). These values also approached those of the semi-optimal scenario. The good performance of this technique attests to its stronger capabilities for identifying reliable negatives, as observed in RQ2. Our multi-stage learning methods rely on the smoothness assumption; that samples close to each other are likely to have the same label, and vice versa. From Figure \ref{fig:embeddings}, we observed this assumption to partially hold. 

Confident learning actually exhibited a decrease in recall compared to the baseline, due to its poor ability to observe label noise in the training data. This may be because confident learning approaches assume the classes to be distinctly separable, which may not be the case. From Figure \ref{fig:embeddings}, we observed relatively small separability between classes. 

The label adjustment strategy appears to impact performance. Pruning labels prior to training (i.e., in Confident Learning and 1-Stage Learning) exhibits slightly worse performance, which is especially apparent for the performance comparison of 1-stage to 2-stage learning. This performance difference is likely due to the lower number of training samples available to pruning methods. 

Knowledge of the noisy labels also serves to significantly increase the number of positive data samples and remediate the class imbalance issue. From Table \ref{tab:data_stats}, we observed that there are a considerably higher number of latent vulnerabilities than reported vulnerabilities. However, simple class rebalancing failed to improve AUC or recall, and is outperformed by the NLL techniques. Hence, we demonstrate that the additional knowledge gained from noisy labels effectively strengthens classifier performance. 

The post-processing technique also performed reasonably well. The performance was comparable to 1-stage learning, despite it applying noise adaptation to the outputs rather than sample selection of the inputs. This may suggest that effectiveness of a method is more heavily dependent on its data assumptions and ability to infer noise, rather than the stage of its application. 

Like in RQ2, we observed low precision for all methods. However, this performance is comparable to existing works \cite{theisen2020better}, so it is not unexpected. Precision could be increased  by adjusting the probabilty classification threshold. However, for reasons discussed in Section 4.5, precision was not the focus of our study, and we hence optimized the classification threshold towards recall whilst minimizing the false positive rate. Furthermore, all NLL techniques were able to achieve a precision higher than the baseline, with a 18.5\% and 11.1\% increase for 1-stage and 2-stage learning, respectively. 

We finally compared whether any of the noise characteristics identified in RQ1 affected the proportion of correct predictions made by the 2-stage learning approach in comparison to the baseline. We investigated whether 2-stage learning correctly classified a different proportion of vulnerabilities for particular CWE types of file locations, as we identified these to be significant factors in RQ1. However, we failed to reject the null hypothesis for each case, using a Chi-square goodness of fit test at the 5\% significance level; i.e., the correct predictions were not significantly different for any type. This indicates that NLL methods failed to discover that noisy labels manifest more commonly for particular SV types or file locations. 

\section{Discussion}
\label{sec:discussion}

Our findings suggest that NLL techniques are able to provide additional knowledge to SVP models that is able to improve their effectiveness. We demonstrate this in RQ3, as NLL techniques outperform SVP baselines, even despite class rebalancing. However, the noise identification component of NLL techniques is far from perfect, as demonstrated through their poor precision, and their inability to distinguish specific label noise characteristics. This leads us to speculate challenges towards the adoption of these techniques, and directions for future improvement. 

\subsection{Difficulties in Adoption}


Firstly, class imbalance is a significant issue prevalent in SV datasets \cite{croft2021data}, that negatively influences prediction capabilities of data-driven models. Although solutions to class imbalance exist, such as rebalancing and reweighting, both of which we have investigated and implemented, they are far from complete. Class imbalanced datasets are similarly ill-suited for many NLL techniques, as many NLL techniques rely on the notion of \textit{prediction confidence}. For example, a commonly observed trait of noise-free samples is that they exhibit small loss \cite{huang2019o2u}, or have high prediction probabilities \cite{reed2014training}. However, for imbalanced datasets the minority class exhibits larger loss and smaller prediction probabilities, as models typically fit more strongly to the majority class. In this sense, the majority of the vulnerable samples in our dataset would be observed as label noise, which is actually the opposite of reality. Class imbalance also prevents the use of probabilistic methods \cite{reis2018probabilistic}, as class weights are already imbalanced for imbalanced learning problems. 

Through RQ1, we demonstrated that label noise is affected by semantic features such as SV type and file functionality. Hence, we are constrained to NLL techniques suitable for detecting semantic label noise. However, we found in RQ3 that the investigated NLL techniques failed to detect these semantic trends. The majority of current NLL research is directed towards instance-independent label noise \cite{song2020learning}, which limited the amount of NLL techniques available for use. Thus, there is a need for more NLL techniques that are better suited towards semantic label noise. 

Secondly, the lack of verifiability of the negative (non-vulnerable) labels causes challenges. Without confident or noise free labels for both classes, we cannot apply more traditional multi-round learning approaches, such as self-training \cite{zhu2005semi}. This factor also prevents us from obtaining a balanced noise-free set, for the purposes of validation and guidance. For instance, bilevel learning \cite{jenni2018deep} uses a noise-free validation set to regularize the overfitting of a model through a bilevel optimization approach. 


Sample selection techniques have been shown to be effective for semantic noise \cite{song2020learning}, and hence have been the focus of analysis for this study. However, these techniques rely on several data assumptions, such as smoothness and separability. These assumptions may not be entirely accurate for our data, as suggested by the poor performance produced from the Confident Learning techniques. The feature representations of our data may also simply be insufficient for NLL techniques. 

\subsection{Future Work}
We motivate the need for significant further research into this area. Overcoming data challenges is vital for the success of SVP and other data-driven tasks. Firstly, researchers ought to work towards overcoming the current challenges we have outlined preventing effective adoption of these techniques. We encourage development of NLL techniques specific to the software engineering and security domains. Furthermore, this study is limited in scope and requires further analysis and extension. 

State-of-the-art SVP methods have moved towards finer granularities \cite{lin2020software}, such as function \cite{li2018vuldeepecker} or statement level \cite{li2021vulnerability}. It is important to investigate how NLL techniques can be extended to support these contexts. Given method-level prediction follows similar labelling processes to file-level prediction, the negative dataset is expected to exhibit similar noise characteristics. However, since method-level prediction more heavily relies on patched lines to localise labels to methods, the verification of individual vulnerable lines may become harder and less scalable. Hence, differences in noise may require consideration of different NLL techniques. Finer granularity methods also typically use DL models, so we need to also investigate the capabilities of these techniques when applied to DL architecture. 

The precision of the NLL techniques can also be greatly improved. Whilst they are able to detect many of the latent vulnerabilities, this comes at a tradeoff to precision. Similarly, all techniques produced low precision scores for RQ3, which limits their actual usability and real-world applications. The techniques were still able to achieve decent AUC values, which indicates that it is important to find an appropriate classification threshold that achieves a desired tradeoff between recall and precision. However, further work is still needed towards improving the precision of these methods. 

Approaches that specifically target noise prone SV types or file locations may also be beneficial. For instance, we found in RQ1 that memory-related SV types, such as buffer overflow or use after free vulnerabilities, had higher latency periods and hence a higher proportion of noisy labels. NLL architecture that is weighted towards detecting noise in memory critical components may yield a performance increase. 

Whilst we have conducted our investigation into the SVP task, automated software security analysis involves a variety of tasks such as SV assessment and prioritization \cite{le2021survey}. Hence, we ought to also investigate how these techniques adapt to a multiclass, rather than binary classification scenario. 

Finally, there are other weak supervision methods besides NLL, that may be useful for this domain. For instance, active learning uses human intervention to provide strong label indicators for the most important data points \cite{zhou2018brief}. Yu et al. \cite{yu2019improving} demonstrated the effectiveness of active learning to assist with vulnerability inspection efficiency. Whilst consideration of these techniques was out of scope for this work, future investigation may be fruitful. 

\subsection{Threats to Validity}
\label{sec:threats}

\textit{Internal Validity.} We acknowledge that our use of NLL techniques may be sub-optimal, as we are one of the first to apply them to this domain and scenario.  Furthermore, as the true noise distribution within the unlabeled set is unknown, we may breach required data assumptions for NLL. Alternatively, our findings may be biased towards particular noise types, as for instance, we only used the top 25 CWE types to search for unreported vulnerabilities. However, our experiments were still able to show that our preliminary application of these techniques can improve SVP performance. Whilst we tuned the appropriate hyperparameters of all baselines and models, this tuning may be sub-optimal. Similarly, we did not tune CodeBERT and instead utilized base static feature extraction. However, our investigation focuses on comparative analysis, so any potential shortcomings would likely affect model performance equally, and hence have minimal impact on our findings. 

\textit{External Validity.} Our results may not sufficiently generalize to other applications or datasets. We only conducted our analysis on the Mozilla Firefox dataset. However, label noise has been widely acknowledged as an issue by many researchers for many SV datasets~\cite{croft2021data}. We plan to investigate how NLL techniques translate to other SV datasets in future. 

\textit{Construct Validity.} Verification of our positive set (vulnerable files) was done manually. Hence, like any other task requiring human raters, there may be inaccuracies or bias. We have attempted to minimize such threats through the use of two raters, and we have reported the inter-rater agreement. 


\section{Conclusion}
\label{sec:conclusion}

We have analysed label noise in security defects and demonstrated its impact on prediction models. The significant difficulty of removing these data quality issues motivated us to investigate solutions that maximized the power of limited label data. Specifically, we proposed to overcome inevitable data issues through noise tolerant learning approaches for automated software security analytics. We applied several noisy label learning techniques taken from the existing literature to software vulnerability prediction baseline models. We observed some noisy label learning methods to exhibit promising performance increases over the baseline, increasing model recall by an average of 18.8\%. However, none of the techniques were able to reach or exceed the performance of the scenario in which we have semi-omniscient knowledge of the label noise. 

Hence, whilst noisy label learning methods have promise for this domain, and can effectively improve performance, there are currently several difficulties preventing their optimal use. We have outlined several key future research directions, and promote further investigation of noise quality and tolerance in the software engineering domain. 


\begin{acks}
We would like to acknowledge Chirag Thakkar for assistance with data collection, and Triet Le for initial discussion and guidance. This work has been supported by the Cyber Security Cooperative Research Centre Limited whose activities are partially funded by the Australian Government’s Cooperative Research Centre Programme. 
\end{acks}

\bibliographystyle{ACM-Reference-Format}
\bibliography{bibfile}

\end{document}